\documentclass[%
 reprint,
superscriptaddress,
 amsmath,amssymb,
 aps,
]{revtex4-1}
\newcommand{\specificthanks}[1]{\@fnsymbol{#1}}

\usepackage{graphicx}
\usepackage{dcolumn}
\usepackage{bm}
\usepackage[colorlinks]{hyperref}
\hypersetup{urlcolor=blue,linkcolor=black,citecolor=black}

\preprint{APS/123-QED}
\begin{document}

\title{An advanced apparatus for the integration of nanophotonics and cold atoms}

\author{J.-B. B\'eguin}\thanks{These authors contributed equally to this work}\affiliation{Norman Bridge Laboratory of Physics, California Institute of Technology, Pasadena, California 91125, USA}
\author{A. P. Burgers}
\altaffiliation[Present address: Department of Electrical Engineering, Princeton University, Princeton, New Jersey 08540, USA]{}
\affiliation{Norman Bridge Laboratory of Physics, California Institute of Technology, Pasadena, California 91125, USA}

\author{\hspace{-1.5mm}$^{,}\hspace{.5mm}^{*}$ X. Luan}
\thanks{These authors contributed equally to this work}
\affiliation{Norman Bridge Laboratory of Physics, California Institute of Technology, Pasadena, California 91125, USA}
\author{Z. Qin}
\thanks{These authors contributed equally to this work}
\affiliation{Norman Bridge Laboratory of Physics, California Institute of Technology, Pasadena, California 91125, USA}
\affiliation{State Key Laboratory of Quantum Optics and Quantum Optics Devices, Institute of Opto-Electronics, Shanxi University, Taiyuan 030006, China}
\author{S. P. Yu}
\homepage[Present address: Time and Frequency Division, NIST, 385 Broadway, Boulder, Colorado 80305, USA]{}
\affiliation{Norman Bridge Laboratory of Physics, California Institute of Technology, Pasadena, California 91125, USA}
\author{H. J. Kimble}
\email{hjkimble@caltech.edu}
\affiliation{Norman Bridge Laboratory of Physics, California Institute of Technology, Pasadena, California 91125, USA}


\date{\today}

\pacs{Valid PACS appear here}

\title{An advanced apparatus for the integration of nanophotonics and cold atoms}




\begin{abstract}
We combine nanophotonics and cold atom research in a new apparatus enabling the delivery of single-atom tweezer arrays in the vicinity of photonic crystal waveguides.
\end{abstract}

\maketitle

\section{Introduction}
There are growing research communities whose goal is to create new paradigms for strong quantum interactions of light and matter by way of single atoms and photons in nanoscopic dielectric lattices~\cite{Kimble:18}. Nanophotonic structures offer the intriguing possibility to control atom-photon interactions by engineering the properties of the medium through which they interact. By now, diverse theoretical opportunities have emerged for atoms and photons in one and two-dimensional nanophotonic lattices, including novel quantum phases and chiral quantum optics~\cite{Kimble:18,Lodahl:17}. By contrast, advances on an experimental front have been slower due to difficulties in suitable nanophotonic device fabrication and the integration of novel devices into the realm of ultracold atoms~\cite{Tiecke:2014,Hood:2016,Thompson:2013,Goban:2014,Burgers:2019}. The significance of this multi-disciplinary research effort relates to diverse applications in Quantum Information Science, including the realization of complex quantum networks, the exploration of quantum many-body physics with atoms and photons, and the investigation of quantum metrology with integrated functionality of nanophotonics and atoms.



\begin{figure}[t]
\centering
\includegraphics[width=1\linewidth]{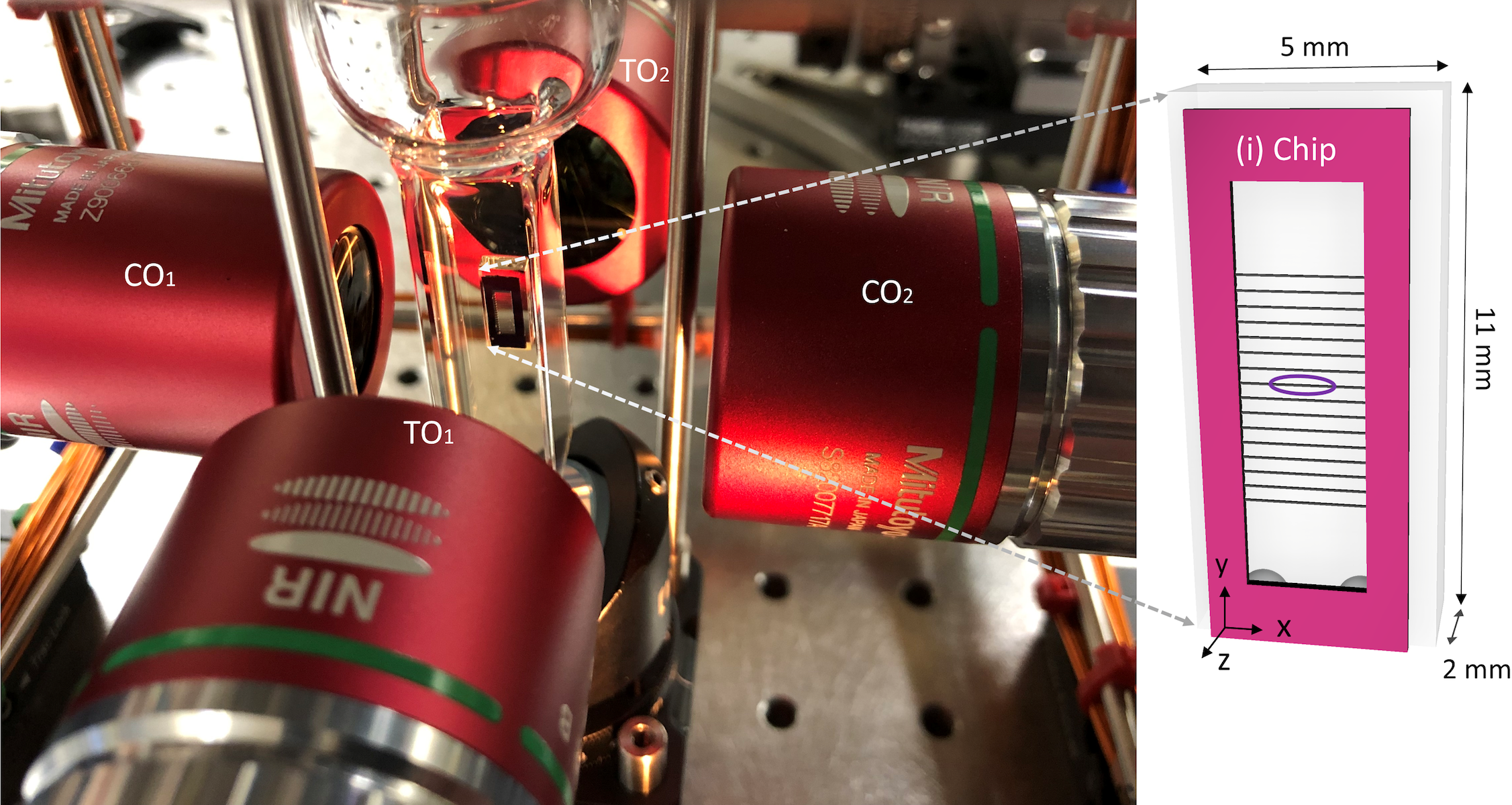}
\vspace{-5mm}
\caption{Advanced apparatus that combines free-space optical coupling to photonic crystal waveguides (PCWs) within the vacuum envelope of an SiO$_2$ cell (i.e., the red-ended coupling objectives CO$_1$ and CO$_2$) and optical tweezers (i.e., the labelled `tweezer objectives' TO$_1$ and TO$_2$) that have large numerical aperture.  Inset: A Si chip with 16 PCWs is bonded to a small SiO$_2$ `table', which is in turn mounted inside the SiO$_2$ cell, with an inner wall spacing of 1 cm. Cold atoms are delivered to the cell along the vertical axis of the vacuum cell.}
\vspace{-5mm}
\label{fig:setup}
\end{figure}

\begin{figure*}[t]
\centering
\includegraphics[width=1\linewidth]{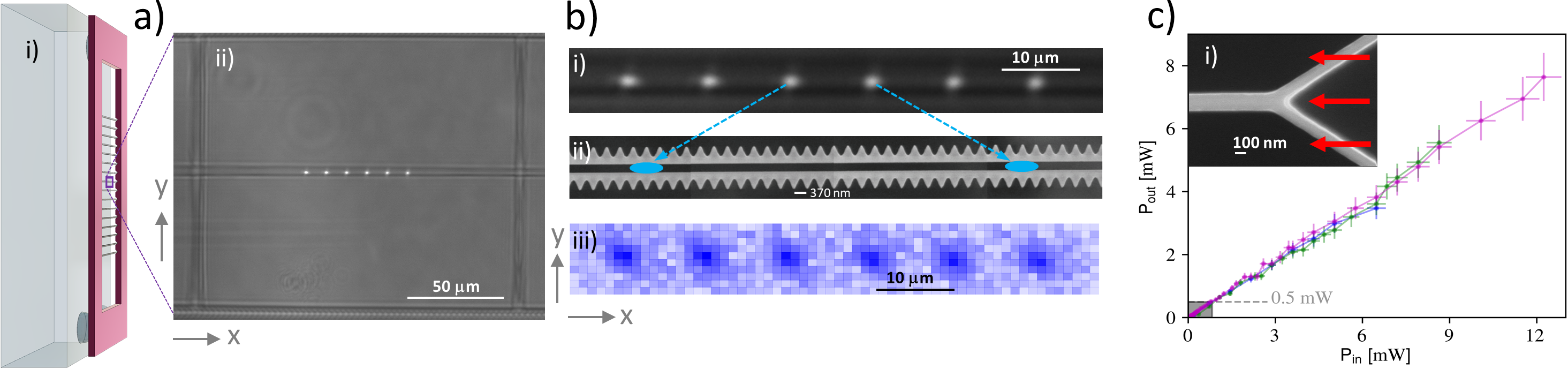}
\vspace{-5mm}
\caption{a) (i)The SiN chip containing the 16 PCWs suspended across a 2 mm $\times$ 7 mm window is silicate bonded to a polished optical table. (ii) The central region of the waveguide containing the PCW is illuminated by six tweezer spots. The PCW is constrained transversely along y by sets of tethers at either side of the image. b)  (i) Zoomed in scattering of the six tweezer spots on the waveguide. The light from the tweezers is reflected from the PCW and imaged through the tweezer objective. (ii) an SEM image of the PCW that reflects the tweezer light. Cyan ellipses indicate the separation of the tweezer spots and approximate the confinement of an atom trapped with energy 0.5 mK (i.e. half the trap depth). (iii) Free-space atomic fluorescence from loading of the six tweezer sites with a 1.26 $\mu$m beam waist for 150 experimental shots 3 mm away from the chip structure. c) Power handling of the free-space coupler (SEM inset (i)) used to launch guided mode light into the PCW with the 0.5 mW limit of previous devices shown~\cite{Hood:2016,Yu:14}.}
\vspace{-5mm}
\label{fig:components}
\end{figure*}

In this Letter, we report the development of an advanced apparatus that overcomes several significant barriers to laboratory progress for cold atoms and nanophotonics. Fig.   \ref{fig:setup}  is a photograph of our `physics cell' that achieves the following set of capabilities. (1) By way of silicate bonding \cite{Veggel:14,Cumming:12}, we stably secure our Silicon chips to the inner walls of small fused silica cells, thereby reducing the volume of the (bake-able) vacuum envelope to ~$\sim 5$ cm$^3$ with unprecedented optical access and lifetime for trapped atoms $\simeq 10-25$ s. This ultra-stable precision method~\cite{Gwo:01} used in gravitational wave detectors can be adapted to solve the practical challenge of cold atoms integration with nano-photonics Silicon chips. (2) The small size of the `physics cell' and its unencumbered optical access allow commercially available long-working distance microscope objectives with large numerical aperture (N.A $\sim$ 0.4 - 0.7) to be configured around the cell for readily achieving single-atom trapping and imaging Fig.~\ref{fig:components}b), atom transport through external mechanical motion of the various objectives as well as simultaneous coupling of free-space laser light into nano-waveguides. (3) With the novel design of photonic crystal waveguides (PCWs) based upon free-space coupling of light to/from our PCWs~\cite{Yuthesis:17}, we have eliminated all optical fibers within the vacuum envelope, achieved nano-waveguide coupling efficiency from free-space laser light up to $80\%$, and increased the power handling capabilities of the PCWs (from $\sim 0.5$ mW to $\sim$10 mW, Fig.~\ref{fig:components}c)). The latter extends roughly 20x beyond the failure power for our previous butt-coupled devices \cite{Yu:14}. Inset of Fig.  \ref{fig:components}c) shows an SEM image of the waveguide free-space coupler. The Y-shape design reduces the dielectric polarizability induced by the propagating light field.
 
In our apparatus, cold atoms enter the physics cell by freely falling along y from a source chamber  $\sim0.5$ m above the science cell. Transverse confinement is provided by a blue-detuned `donut' beam. In the physics chamber atoms are stopped and cooled by Polarization Gradient Cooling (PGC) to a volume of $\sim (200 \mu m)^3$ with temperature $\simeq 20$ $\mu$K. Next, individual Cesium atoms from the PGC cloud are loaded into a linear array of optical tweezers operated at a magic wavelength (936 nm) for Cs deep in a regime of `collisional blockade' \cite{Ye:08,Schlosser:02}. This results in loading of either 0 or 1 atom with approximately 50\% probability into each tweezer spot. Fig.  \ref{fig:components}b(i) displays the reflection of multiple tweezer spots off the PCW. An example of atoms loaded into free-space tweezers far from the PCW is shown in Fig.~\ref{fig:components}b(iii). The optical tweezers are formed by objective TO$_1$ with NA = 0.4, shown in Fig. \ref{fig:setup}. We measured focal spot waists within the evacuated cell of $w_0 \simeq 1.26 \pm 0.15 \mu$m. Sub-micron waists have been achieved with a higher numerical aperture objective (NA = 0.67). To minimize the impact of light scattering from the chip during trap loading, the tweezer array is loaded approximately 3 mm away from the surface of the Silicon chip. Due to the relatively long trap storage lifetime, transport of atoms trapped in the tweezer array to near the surface of the Silicon chip along the PCW is accomplished over a programmable interval $0.02 <\Delta t < 0.1$ s by mounting TO$_1$ on a precision linear translation stage along $z$ (model Physik Instrumente V-522, $20$ nm unidirectional repeatability). Fig. ~\ref{fig:components}b(ii) shows an SEM of the underlying alligator photonic crystal waveguide (APCW) representing the target destinations. Measurements that quantify the survival probability $P_s$ for tweezer-trapped atoms to be moved from the loading zone to the PCW and back suggest $P_s \sim 0.9$.

Our goal in this effort is to achieve strong quantum interactions of light and matter by way of single atoms and photons in nanoscopic dielectric lattices that are assembled deterministically with arrays of single-atom tweezers to enable the exciting physics described in~\cite{Kimble:18}. This effort is in the spirit of recent worldwide advances with free-space tweezer arrays~\cite{Endres:2016,Barredo:2016}. Our effort adds the difficulty of deterministic assembly of such atomic arrays near the surfaces of 1-D and 2-D PCWs \cite{Yu:19}, which the advances described in this Letter address.

\noindent\textbf{Funding Information.} We acknowledge support from the following grants and organizations: ONR Grant No. N000141612399, ONR MURI Quantum Opto-Mechanics with Atoms and Nanostructured Diamond Grant No. N000141512761, AFOSR MURI Photonic Quantum Matter Grant No. FA95501610323, and NSF Grant No. PHY1205729, as well as the Caltech KNI.

\noindent\textbf{Acknowledgments.} The authors thank John Hall, Jun Ye, Norma Robertson, and Keith Hulme for important discussions.

\noindent\textbf{Disclosures.} The authors declare no conflicts of interest.





\bibliography{optica_draft_arxiv_v1}

\end{document}